\documentclass[12pt]{article}
\usepackage{graphicx}
\usepackage{amsmath,amssymb,amsfonts}

\begin{document}

\title{Gravitational radiation from a rotating magnetic dipole}

\author{S. Hacyan
}

\renewcommand{\theequation}{\arabic{section}.\arabic{equation}}

\maketitle
\begin{center}

{\it  Instituto de F\'{\i}sica,} {\it Universidad Nacional Aut\'onoma de M\'exico,}

{\it A. P. 20-364, Cd. de M\'exico, 01000, Mexico.}

\end{center}
\vskip0.5cm

\begin{abstract}

The gravitational radiation emitted by a rotating magnetic dipole is calculated. Formulas for the polarization
amplitudes and the radiated power are obtained in closed forms. A comparison is made with other sources of
gravitational and electromagnetic radiation, particularly neutron stars with extremely powerful magnetic
fields.

\end{abstract}



\section{Introduction}

Gravitational radiation is an important source of energy in many astrophysical phenomena. A neutron star, for
instance, radiates both electromagnetic \cite{pac,hac} and gravitational waves \cite{bona,cut,jones}; the main
sources of radiation are the interior of the star (behaving as a magnetized fluid), the external magnetic field
and the corotating magnetosphere.

In the present article, we study the gravitational waves (GWs) generated by one of these possible sources: a
rotating magnetic dipole. In Section 2, the radiated energy and the polarization amplitudes of the GWs are
calculated using the quadrupole formula and considering the electromagnetic field in the near zone of the dipole,
where most of the energy of the field  is located. The results are discussed in Section 3 and compared with other sources
of radiation, electromagnetic or gravitational, with a focus on stars such as magnetars\cite{magnetar} that  possess extremely powerful
magnetic fields.

\section{Near zone}

Our starting point is the formula for the metric $h^{TT}_{ij}$ in the TT gauge  (see Maggiore \cite{magg} for notation and details),
\begin{equation}
h^{TT}_{ij} = \frac{1}{r}~\frac{2G}{c^4} \Lambda_{ij,kl}({\bf \hat{n}}) \ddot{Q}^{kl} (t-r/c) ,\label{TT}
\end{equation}
where dots represent derivation with respect to time $t$,
\begin{equation}
\Lambda_{ij,kl}({\bf \hat{n}})= P_{ik}P_{jl}- \frac{1}{2}P_{ij}P_{kl}, \label{lambda}
\end{equation}
and
\begin{equation}
P_{ij} ({\bf \hat{n}}) =\delta_{ij}- \hat{n}_i\hat{n}_j
\end{equation}
is the projection tensor with respect to the unit vector ${\bf \hat{n}}$. The quadrupole is defined as
$Q^{ij}= M^{ij} - \frac{1}{3} \delta^{ij}M^{nn}$ in terms of
\begin{equation}
M^{ij}= \frac{1}{c^2}\int T^{00} x^i x^j ~dV,\label{dV}
\end{equation}
where $T^{00}$ is the $00$ component of the energy-momentum tensor.

The energy radiated in the form of GWs
in the direction of ${\bf \hat{n}}$ is
\begin{equation}
\frac{dE}{d\Omega} = \frac{r^2c^3}{32 \pi G}  \int_{-\infty}^{\infty} dt ~ \dot{h}_{ij}^{TT}
\dot{h}_{ij}^{TT}.\label{2}
\end{equation}

\subsection{Magnetic dipole}

Consider the field of a magnetic dipole of magnitude $m$. In the near zone, the electric field can be neglected
and the energy density of the magnetic field  is
\begin{equation}
T_{00} = \frac{1}{8 \pi} \frac{m^2}{r^6} \Big(1+3 ({\bf \hat{r}} \cdot {\bf \hat{u}}(t)) ^2\Big),\label{T}
\end{equation}
where ${\bf \hat{u}}(t)$ is the unit vector in the direction of the dipole, and ${\bf \hat{r}}$ is a unit radial
vector. Further corrections to the electromagnetic field are of order $\omega r/c$ with respect to $B_i$ (where $\omega$ is
the rotation frequency of the dipole). For neutron stars of radius $R \sim 10$ km, the
approximation is valid for $\omega \ll c/R \sim 3 \times 10^4~ {\rm s}^{-1}$.

It follows with some straightforward algebra that
\begin{equation}
Q^{ij} (t) = \frac{m^2}{5Rc^2} \Big(\hat{u}^i(t) \hat{u}^j(t) -\frac{1}{3} \delta^{ij}\Big),
\end{equation}
where $R$ is a lower cut-off that can be identified with the radius of the star. It is understood that the volume
integral \eqref{dV} covers the region $r \geq R$.

The metric $h_{ij}^{TT}$ follows from the above formula and Eq. \eqref{TT}:
\begin{equation}
h_{ij}^{TT}= \frac{1}{r}~\frac{2Gm^2}{5 Rc^6} ~\Lambda _{ij,kl} ~ \frac{d^2}{dt^2} \Big(\hat{u}^k (t_r) ~\hat{u}^l
(t_r)\Big),\label{htt}
\end{equation}
where  $t_r=t-r/c$.

Let us now take a coordinate system in which the rotation axis of the dipole is in the $z$ direction. Thus
\begin{equation}
{\bf \hat{u}}(t) =( u_{\bot}\cos (\omega t),  u_{\bot}\sin (\omega t), u_{\parallel}),
\end{equation}
where $u_{\parallel}$ is the constant component of ${\bf \hat{u}}(t)$ along the rotation axis and $u_{\bot}^2 =1 -
u_{\parallel}^2.$ In this same system of coordinates we can
define the three orthonormal vectors
\begin{eqnarray}
{\bf \hat{n}} &=& (\sin \theta \cos \phi,~ \sin \theta \sin \phi,~\cos \theta), \\ \nonumber
 \hat{\boldsymbol{\theta}} &=& (\cos \theta
\cos \phi, ~\cos \theta \sin \phi,~-\sin \theta),\\ \nonumber \hat{\boldsymbol{\phi}}  &=& (-\sin \phi, ~\cos
\phi,~0), \\ \nonumber
\end{eqnarray}
together with the useful formulas
\begin{equation}
\hat{\theta}_i \hat{\theta}_j \Lambda_{ij,kl}= \hat{\theta}_k \hat{\theta}_l - \frac{1}{2}P_{kl} ,\nonumber
\end{equation}\begin{equation}
\hat{\phi}_i \hat{\phi}_j \Lambda_{ij,kl}= \hat{\phi}_k \hat{\phi}_l - \frac{1}{2}P_{kl}, \nonumber
\end{equation}
\begin{equation}
 \hat{\phi}_i \hat{\theta}_j \Lambda_{ij,kl}= \hat{\phi}_k \hat{\theta}_l .\label{bla}
\end{equation}

\subsection{Metric}\label{met}

The two metric potentials of the GW can be calculated from Eq. \eqref{htt} and the formulas \eqref{bla}. The result is
$$
h_+ \equiv h_{ij}^{TT} \hat{\phi}_i \hat{\phi}_j = -h_{ij}^{TT} \hat{\theta}_i \hat{\theta}_j=\frac{1}{r}
~\frac{Gm^2}{5Rc^6}\frac{d^2}{dt^2} ( u_{\phi}^2-u_{\theta}^2)
$$
\begin{equation}
h_{\times} \equiv -h_{ij}^{TT} \hat{\theta}_i \hat{\phi}_j = -\frac{1}{r} ~\frac{2Gm^2}{5Rc^6} \frac{d^2}{dt^2}
(u_{\theta}u_{\phi}),\label{otro}
\end{equation}
where $u_{\theta}= {\bf \hat{u}} \cdot \hat{\boldsymbol{\theta}}$ and $u_{\phi}= {\bf \hat{u}} \cdot
\hat{\boldsymbol{\phi}}$, and  $({\bf \hat{u}} \cdot {\bf \hat{n}})^2+ u_{\theta}^2 +u_{\phi}^2 =1$. The above
two formulas can be written  as
\begin{equation}
h_+ + i h_{\times} = \frac{1}{r}~\frac{Gm^2}{5Rc^6}\frac{d^2}{dt^2} (u_{\phi} -i u_{\theta})^2.
\end{equation}
Explicitly
$$
u_{\phi}=u_{\bot} \sin (\omega t')
$$
\begin{equation}
u_{\theta}= u_{\bot} \cos \theta \cos (\omega t')-u_{\parallel} \sin \theta,
\end{equation}
with $\omega t' = \omega t_r - \phi $, from where it follows that
$$
h_+ + i h_{\times} = \frac{1}{r}~\frac{Gm^2}{5Rc^6} \omega^2 u_{\bot} \Big\{ 2u_{\bot} \Big[(1+\cos^2 \theta)~\cos
(2\omega t')+2 i \cos \theta ~\sin (2 \omega t')\Big]
$$
\begin{equation}
-2 u_{\|} \sin(\theta)\Big[~\cos(\theta) \cos (\omega t')
 + i   \sin (\omega t')\Big] \Big\}.\label{hh}
\end{equation}

Accordingly, the spectrum of the GW has two lines, one at $\omega$ corresponding to the $u_{\|}$ component, and
one at $2\omega$ corresponding to the $u_{\bot}$ component (only the latter is present for a GW propagating along
the rotation axis). The amplitude of the wave is of order $Gm^2\omega^2 u_{\bot}/(Rc^6r)$.

\subsection{Radiated energy}

The radiated power can be calculated noticing that $ \dot{h}_{ij}^{TT} \dot{h}_{ij}^{TT}= |\dot{h}_+ +i
\dot{h}_{\times}|^2$. The energy radiated per unit time follows from Eq. \eqref{2} performing the integration over
one period $T=2\pi/\omega$ and dividing by $T$. The result is
\begin{equation}
\frac{dP}{d\Omega}= \frac{G m^4}{ 200 R^2 c^9}u_{\bot}^2\omega^6  \Big[1 - \cos^4 \theta + u_{\bot}^2 (5 \cos^4
\theta +24\cos^2 \theta +3) \Big].\label{ponz}
\end{equation}
Finally, an integration over solid angles yields the total power radiated:
\begin{equation}
P= \frac{\pi G m^4}{75 R^2  c^9}u_{\bot}^2 \omega^6 (1 +18 u_{\bot}^2).\label{Pnz}
\end{equation}

\section{Comparisons and conclusions}

For the dipole field, we can set $m=B_0 R^3$, where $B_0$ is the average strength of the magnetic field at the
surface of the star. If $B_0 \sim 10^{12}$G and $\omega \sim 1~{\rm s}^{-1}$, the power radiated in the form of
gravitational radiation is
$$
P \sim 10^6 \Big(\frac{B_0}{10^{12} {\rm G}}\Big)^4 \Big(\frac{R}{10~ {\rm km}} \Big)^{10} (\omega ~ {\rm s})^6
u_{\bot}^2~ {\rm ergs/s}
$$
according to  formula \eqref{Pnz}. Of course, for average pulsars, this is many orders of magnitude below the
power emitted in the form of electromagnetic waves, which is typically $10^{28}$ ergs/s \cite{pac}. Nevertheless,
for a millisecond magnetar with $B_0 \sim 10^{14}$ G, the power of the  GWs could be of the order of $10^{32}$
ergs/s.

It is also instructive to compare our results with those obtained by Bonazzola and Gourgoulhon \cite{bona} for the
emission of GWs from the interior of a rotating neutron star. These authors obtained a value for the amplitudes of
GWs
\begin{equation}
|h_{+}+i h_{\times}|\sim \frac{1}{r}~\frac{4G}{c^4} I \epsilon \omega^2 ,
\end{equation}
where $I$ is the moment of inertia and $\epsilon$ is the ellipticity of the star; typical values of these
parameters are $\epsilon \sim 10^{-6}$ or smaller, and  $I \sim 10^{45}$ g cm$^2$. If we compare their result with
our Eq. \eqref{hh}, we see that the amplitudes of the GWs produced by the rotating fluid are larger by a factor
$$10^{13} \epsilon~ (B_0  /10^{12}{\rm G })^{-2} $$
than the amplitudes of GWs produced by the rotation of the magnetic field. Thus, for a usual neutron star with
$B_0 \sim 10^{12}$ G, the contribution of the rotating dipole is comparatively negligible. However, it is not
negligible for magnetars having fields $B_0 \sim 10^{14}$ G \cite{magnetar} and rather small deformations
$\epsilon < 10^{-6}$.

In conclusion, the external magnetic field of a magnetar can make a significant correction to the gravitational
radiation produced by the internal magnetized fluid.

\end{document}